\documentclass[prl,twocolumn,showpacs,nofootinbib]{revtex4-1}
\usepackage{amsmath}    % need for subequations
\usepackage{amssymb}
\usepackage{graphicx}% Include figure files
\usepackage{graphics}
\usepackage{epsfig}
\usepackage{verbatim}   % useful for program listings
\usepackage[subfigure]{graphfig}
\usepackage{epsfig}
\usepackage{epstopdf}
\usepackage{dcolumn}% Align table columns on decimal point
\usepackage{color}
\usepackage{CJK}

\usepackage[normalem]{ulem}% for \sout

\def\be{\begin{equation}}
\def\ee{\end{equation}}
\def\bea{\begin{eqnarray}}
\def\eea{\end{eqnarray}}
\newcommand{\Tr}{\mathrm {Tr}}

\newcommand{\unit}{1\!\!1}

\definecolor{green}{rgb}{0,.5,0}

\begin{document}

\begin{CJK*}{UTF8}{} % Use default fonts from CJK (see below)
\CJKfamily{gkai}

\title{\vspace{1.0in} {\bf Strange and Charm Quark Spins from Anomalous Ward Identity}}

\author{Ming Gong (宫明)$^{1}$, Yi-Bo Yang (杨一玻)$^{2}$, Jian Liang (梁剑)$^{2}$, Andrei Alexandru$^{3}$, Terrence Draper$^{2}$, \CJKfamily{bkai}and Keh-Fei Liu (劉克非)$^{2}$ }
\collaboration{$\chi$QCD Collaboration}
\affiliation{
$^{1}$\mbox{Institute of High Energy Physics, Chinese Academy of Science, Beijing 100049, China}\\
$^{2}$\mbox{Department of Physics and Astronomy, University of Kentucky, Lexington, KY 40506, USA}\\
$^{3}$\mbox{Department of Physics, The George Washington University, Washington, DC 20052, USA}\\
}


\begin{abstract}
We present a calculation of the strange and charm quark contributions to the nucleon spin from the
anomalous Ward identity (AWI). It is performed with overlap valence quarks on 2+1-flavor domain-wall fermion gauge configurations on  a $24^3 \times 64$ lattice with the light sea mass at $m_{\pi} = 330$ MeV. To satisfy the AWI, the overlap fermion for the pseudoscalar density and  the overlap Dirac operator for the topological density, which do not have multiplicative renormalization, are used to normalize the form factor of the local axial-vector current at finite $q^2$. 
For the charm quark, we find that the negative pseudoscalar term almost cancels the positive topological term. For the strange quark, the pseudoscalar term is less negative than that of the charm.
By imposing the AWI, the strange $g_A(q^2)$ at $q^2 =0$ is obtained by a global fit of the pseudoscalar and the topological form factors, together with $g_A(q^2)$ and the induced pseudoscalar form factor $h_A(q^2)$ at finite $q^2$. The chiral extrapolation to the physical pion mass gives $\Delta s + \Delta {\bar{s}} = -0.0403(44)(78)$.
\end{abstract}

\maketitle

\end{CJK*}
%\section{Introduction}

The quark spin content of the nucleon was found to be much smaller than that expected from the quark 
model by the polarized deep inelastic lepton-nucleon scattering experiments and the 
recent global analysis reveals that the total quark spin contributes only $\sim 25\%$ to the proton 
spin~\cite{deFlorian:2009vb}.

In an attempt to understand the smallness of the quark spin contribution from first principles,
several lattice QCD calculations~\cite{Dong:1995rx,Fukugita:1994fh} have been carried out since 1995 with the quenched approximation or with heavy dynamical fermions~\cite{Gusken:1999as}.
The most challenging part of the lattice calculation is that of the disconnected insertion of the nucleon three-point functions due to the quark loops. \ Recently,\ the strange quark spin $\Delta s + \Delta {\bar{s}}$ has been calculated
with the axial-vector current on light dynamical fermion configurations~\cite{QCDSF:2011aa,Babich:2010at,Engelhardt:2012gd,Abdel-Rehim:2013wlz,Chambers:2015bka} and it is found to be in the range from $-0.02$ to $-0.03$.
This is about 4 to 5 times smaller in magnitude than that from a global fit of DIS 
which gives $\Delta s + \Delta {\bar{s}} \approx - 0.11$~\cite{deFlorian:2009vb} and a most recent analysis~\cite{Leader:2014uua} including
the JLab CLAS high precision data which finds it to be $-0.106 (23)$~\cite{Stamenov+Leader2015}.

Such a discrepancy between the global fit of experiments and the lattice calculation 
of the quark spin from the axial-vector current is unsettling. It was emphasized some time ago that it is essential that a lattice calculation
of the flavor-singlet axial-vector current be able to accommodate the triangle anomaly~\cite{Karsten:1980wd,Lagae:1994bv}.
It was specifically suggested~\cite{Karsten:1980wd} to calculate the triangle anomaly from the VVA vertex and take it as the normalization condition for the axial-vector current in order to determine the normalization factor $\kappa_A$ on the lattice.  To address the discrepancy of the strange quark spin,  we shall use the anomalous Ward identity (AWI) to provide the normalization and renormalization conditions to calculate the strange and charm quark spins in this work.

The anomalous Ward identity (AWI) is usually referred to the flavor-singlet axial current $A_{\mu}^0 = A_{\mu}^u + A_{\mu}^d + A_{\mu}^s$ in the flavor $SU(3)$ basis where there is a $U(1)$ anomaly term in the divergence of $A_{\mu}^0$. However, the
flavor $SU(3)$ is a global symmetry, AWI is satisfied for each flavor in the flavor basis through linear combinations of the
flavor-octet axial current $A_{\mu}^8 = A_{\mu}^u + A_{\mu}^d  - 2A_{\mu}^s$  and isovector axial current
$A_{\mu}^0 = A_{\mu}^u - A_{\mu}^d$. For the case of the strange quark, its AWI can be obtained from the AWI for the
$A_{\mu}^0$ and the WI for $A_{\mu}^8$ (N.B. there is no anomaly term in the WI for $A_{\mu}^8$) through the
combination $A_{\mu}^s \equiv \frac{1}{3}(A_{\mu}^0 - A_{\mu}^8)$. Alternatively, the AWI can be derived for the
strange by considering the infinitesimal local chiral transformation $\psi \rightarrow \psi + \delta_A \psi$ where
$\delta_A \psi =  i\epsilon(x) \gamma_5 T \psi$ with the $3 \times 3$ matrix in flavor space \mbox{$T = \left( \begin{array}{ccc} 0 & 0 & 0 \\ 0 &0&0 \\ 0&0&1 \end{array} \right) = \frac{1}{3} (\unit - \sqrt{3}\lambda^8)$}, where $\lambda^8$ is the 8th $SU(3)$ generator,  gives a chiral transformation only for the strange. 

For the overlap fermion~\cite{Neuberger:1997fp} which has chiral symmetry on the lattice via the Ginsparg-Wilson relation, the conserved flavor-singlet axial current is derived~\cite{Hasenfratz:2002rp}. Following the derivation with the above definition for the matrix $T$ for the chiral transformation, it is straightforward to show that the following identity is satisfied for the strange axial-vector current.
\begin{equation}\label{awi}
\Bigg \langle i \frac{\delta_A^s \mathcal{O}}{\delta \epsilon(x)} \Bigg \rangle - \langle \mathcal{O}\partial_{\mu}^*  A_{\mu, cons}^s  (x)\rangle + 2m_s \langle \mathcal{O}\, P^s (x)\rangle - 2i \langle \mathcal{O}\, q(x) \rangle =0,
\end{equation}
where $\partial_{\mu}^*$ is the forward derivative. The expression of the conserved current for the strange quark $A_{\mu, con}^s$ is
given in Ref.~\cite{Hasenfratz:2002rp} which involves a non-local kernel which is more involved to implement numerically than the local current. In this work we shall replace it with the local current 
$A_{\mu}^s =  \overline{s} i \gamma_{\mu}\gamma_5 (1-\frac{1}{2}D_{ov}) s$
where $D_{ov}$ is the massless overlap operator which is exponentially local with a fall-off of one lattice 
spacing~ \cite{Draper:2005mh}.
The topological charge $q(x) = \Tr \,\gamma_5 ( \frac{1}{2}D_{ov}(x,x) \!-\!1)$ is derived in the Jacobian factor from the fermion determinant under the chiral transformation which is 
equal to  $\frac{1}{16 \pi^2} tr_c G_{\mu\nu} \tilde{G}_{\mu\nu}(x)$ in the continuum~\cite{Kikukawa:1998pd}, i.e. 
\begin{eqnarray}  \label{top-charge}
\!\!q(x)\! =\!  \Tr \,\gamma_5 ( \frac{1}{2}D_{ov}(x,x) \!-\!1) 
 {}_{\stackrel{\longrightarrow}{a \rightarrow 0}} \frac{1}{16 \pi^2}tr_c \, G_{\mu\nu} 
\tilde{G}_{\mu\nu}(x),
\end{eqnarray}
where $\Tr$ is the trace over both spin and color, while $tr_c$ is the trace over color. For the strange quark spin, we shall consider the $\mathcal{O}$ in Eq.~(\ref{awi}) to be the nucleon propagator, i.e.
\begin{equation}
\mathcal{O} = \Tr [\Gamma_3 \sum_{\vec{z}} e^{-i \vec{p'}\cdot \vec{z}}\chi_(z,t) \sum_{\vec{y}} e^{-i \vec{p}\cdot \vec{y}}
\overline{\chi}_(y,0)]
\end{equation}
where $\chi$ is the commonly used proton interpolation operator which involves two $u$ and one $d$ fields
\begin{equation}
 \chi_\gamma(x) = \epsilon_{abc}\,\psi\,^{\mbox{\scriptsize T} (u)a}_\alpha(x)\,
             (C \gamma_5)_{\alpha\beta} \, \psi^{(d)b}_\beta(x)\, \psi^{(u)c}_\gamma(x), 
\end{equation}
where the Latin letters denotes the color index and the Greek letters denotes the Dirac index and $C = \gamma_2 \gamma_4$ for
the Pauli-Sakurai representation that we adopt for the $\gamma$ matrices. 
In this case, the first term in Eq.~(\ref{awi}) vanishes, since $\mathcal{O}$ does not involve strange quarks and hence 
no $\epsilon (x)$ dependence.  

Following the standard calculation of off-forward  nucleon matrix element~ \cite{Liu:1994dr,Deka:2013zha}, one considers the appropriate combination of the three-point function with the momentum projection of the current $\vec{q} = \vec{p'} - \vec{p}$ and the two-point functions to remove the kinematic dependence and take the time separation between the nucleon source and the current insertion, likewise between the nucleon sink and the current insertion, one arrives at the following unrenormalized AWI in nucleon matrix element for the strange quark
\begin{eqnarray}    \label{nawi}
&&\langle p' s|\partial_{\mu}\kappa_A A_{\mu}^s(q)|p\, s\rangle \nonumber \\
&& = \langle p' s|2m_s P^s(q)|p\, s\rangle - \langle p'  s|2i q(q)|p\, s\rangle
\end{eqnarray}
where $|p\, s\rangle$ is the nucleon state with momentum $\vec{p}$ and spin $s$. As we mentioned above that we shall
replace the conserved axial-vector current $A_{\mu, cons}^s$ with the local one 
$A_{\mu}^s =  \overline{s} i \gamma_{\mu}\gamma_5 (1-\frac{1}{2}D_{ov}) s$. To compensate for the replacement,
a normalization factor $\kappa_A$ is introduced to make the AWI satisfied at finite cutoff. This is the only normalization factor needed
since the pseudoscalar density $P^s$ and the topological charge are the same as those in Eq.~(\ref{awi}) (N.B. In the case of
the disconnected insertion for the strange quark, the pseudoscalar density contributes through the quark loop. In this case,
the $P^s$ takes the form \mbox{$P^s= \overline{s} i \gamma_5(1-\frac{1}{2}D_{ov}) s$}).
This lattice normalization factor is analogous to that introduced to make the chiral Ward identity satisfied for the local non-singlet
axial-vector current. In the literature, it is usually denoted as $Z_A$ which is actually a finite renormalization with no logarithmic 
scale $\mu$ dependence. Following Ref.~\cite{Luscher:1996jn}, we shall call it lattice normalization. Unlike the vector current and non-singlet axial current, the flavor-singlet axial current has, in addition, a renormalization with anomalous dimension. We thus consider the renormalization on top of normalization as is done for the energy-momentum tensor in 
Ref.~\cite{Deka:2013zha}. We will discuss the renormalization after we define the strange quark spin first.

The normalized strange quark spin in the nucleon 
$g_A^{s(N)} \equiv \kappa_A g_A^s = \Delta s + \Delta {\bar{s}}$, where $g_A^s$ is the bare forward matrix element from the local axial-vector current
\begin{equation}
g_A^s s_{\mu}= \frac{\langle p\, s  |  A_{\mu}^s| p\, s \rangle}{\langle p, s  |  p\, s \rangle},
\end{equation}
can be obtained by evaluating the right-hand-side of the
AWI in Eq.~(\ref{awi}) between the nucleon states in the forward limit, i.e.
\begin{eqnarray} \label{eq:awi}
g_A^{s(N)} &=& \lim_{|\vec{q}| \rightarrow 0} \frac{i | \vec{s} |}
{\vec{q} \cdot \vec{s}} \frac{\langle p^\prime s  |  2  m_s P^s\!-\! 2 i q \, | p\, s \rangle}{\langle p^\prime s  |  p\, s \rangle} \nonumber \\
&=& \frac{m_s}{m_N} g_P^s(0) + g_G(0),
\end{eqnarray}
where  $g_P(0)$ and $g_G(0)$ are form factors at $q^2=0$ as defined in Eq.~(\ref{eq:awi}). The normalized charm spin $g_A^{c (N)}= \Delta c + \Delta {\bar{c}}$ is similarly defined. 
In this case, one can, in principle, calculate $g_P(q^2)$ and $g_G(q^2)$ at finite $q^2$ and extrapolate them to the $q^2 \rightarrow 0$ limit and this approach has been studied before~\cite{Mandula:1990ce,Altmeyer:1992nt}; however, the pseudoscalar density term was not included. Despite the fact that there is no massless pseudoscalar pole in the flavor-singlet case, it is shown that the contribution of the pseudoscalar density does not vanish at the massless limit~\cite{Liu:1991ni,Liu:1995kb}. Furthermore, there is a pion pole
in the disconnected insertion of $g_P(q^2)$ to cancel that in the connected insertion to lead to 
$\eta$ and $\eta'$ poles~\cite{Liu:1991ni,Liu:1995kb}. Thus, the $g_P(q^2)$ and $g_G(q^2)$ form factors at 
small $q^2$ of the order of $m_{\pi}^2$ are essential for a reliable $q^2 \rightarrow 0$ extrapolation. Since the smallest 
$- q^2 = 0.21 \,\rm{GeV}^2$ is larger than $m_{\pi}^2 = 0.11\, \rm{GeV}^2$ on the lattice we work on, a naive extrapolation of
$q^2 \rightarrow 0$ in Eq.~(\ref{eq:awi}) may lead to a wrong result. To alleviate this concern, we shall consider instead, in this work, matching the $g_A(q^2)$ and the induced pseudoscalar form factor $h_A(q^2)$ from the left side of Eq.~(\ref{nawi}) and $g_P (q^2)$ and $g_G(q^2)$ from the right side at finite $q^2$ to
determine the normalization constant $\kappa_A$ as will be discussed later.

As far as renormalization is concerned, we note that in the continuum calculation~\cite{Espriu:1982bw},
the renormalization constants of the quark mass and the pseudoscalar density cancel, i.e. $Z_m\,Z_P = 1$, and the
renormalized topological charge term $-2iq$ has a mixing with the divergence of the axial current at one-loop in the form
$\lambda\,  \partial_{\mu} A_{\mu}^{0}$ where $A_{\mu}^{0}$ is the flavor-singlet axial current and 
$\lambda = - (\frac{g_0^2}{4\pi^2})^2 \frac{3}{8} C_2 (R) \frac{1}{\epsilon}$
with one of the $g_0^2$ coming from the definition of the topological charge. 
On the 
other hand, the renormalization of the divergence of axial-vector current occurs at the two-loop level  involving a  quark loop in the disconnected insertion which gives~\cite{Espriu:1982bw}the divergence of the renormalized  strange axial-vector  
\begin{equation}
\partial_{\mu}A_{\mu}^{s (R)} =  \partial_{\mu} A_{\mu}^{s} 
+ \lambda \partial_{\mu}A_{\mu}^0
\end{equation}
In the present work, we adopt the overlap fermion for the lattice calculation where 
$Z_m\,Z_P = 1$ and there is no multiplicative renormalization of the topological charge defined by the overlap operator in Eq.~(\ref{top-charge}). After two-loop matching from the
lattice to the $\overline{MS}$ scheme,  the renormalized and normalized AWI equation at the 
scale $\mu$ is therefore
\begin{eqnarray}\label{rawi}
&&\langle p' s|\partial_{\mu} \kappa_A  A_{\mu}^s + \gamma (\ln(\mu^2 a^2) +f)    \partial_{\mu}  A_{\mu}^0 |p\, s\rangle  \nonumber \\
 &&=  \langle p' s|  2m_s P^s  - 2i  q  + \gamma (\ln(\mu^2 a^2) +f')  \partial_{\mu}  A_{\mu}^0|p\, s\rangle ,
\end{eqnarray}
where $\gamma = - (\frac{g_0^2}{4\pi^2})^2 \frac{3}{8}C_2 (R)$ is the anomalous dimension. We see that, modulo the possible different finite terms $f$ and $f'$ in the renormalization of $A_{\mu}^0$ and the topological charge $q$~\cite{Capitani:2002mp}, 
the anomalous dimension term on the l.h.s is the same as that on 
the r.h.s.~\cite{Espriu:1982bw}.
Thus, the two-loop renormalized AWI is the same as the {\it unrenormalized} AWI in Eq.~(\ref{nawi}).

Two loop renormalization on the lattice is quite involved, we plan to carry out the
calculation of the lattice matching to the $\overline{MS}$ scheme non-perturbatively as is recently done in Ref.~\cite{Chambers:2015bka}. For the present work, we shall give an
estimate of the renormalization correction. From the left side of Eq.~(\ref{rawi}), one 
finds the renormalized $g_A^{s (R)}$
\begin{equation}   \label{delta_ga}
g_A^{s(R)}= g_A^{s(N)} + \delta g_A^s,
\end{equation}
where $g_A^{s(N)} = \kappa_A g_A^s$ is the normalized $g_A^s$ and
\begin{equation}
\delta g_A^s =  \gamma (\ln(\mu^2 a^2) +f) g_A^{0}
\end{equation}
where $g_A^{0}$ is the flavor-singlet $g_A$.

To estimate the size of $\delta g_A^{s (R)}$ for renormalization and matching to the 
$\overline{MS}$ scheme at $\mu = 2$ GeV, we note that $g_0^2 = 2.82$ for the Iwasaki gauge action for DWF configurations, the lattice spacing $a^{-1} = 1.73$ GeV, and we
assume $f$ to be $10$. In this case, \mbox{$\gamma (\ln(\mu^2 a^2) +f) \sim 0.079 $.} Taking the experimental value of  $g_A^0= 0.25$ from experiments~\cite{deFlorian:2009vb}, we obtain $|\delta g_A^s| \sim 0.0066$. We shall take this as a part of the systematic error.

%\section{Technique details}

    We use the overlap fermion for the valence quarks in the nucleon propagator as well as for the quark loops on $2+1$ flavor domain-wall fermion (DWF) configurations on a $24^3 \times 64$ lattice with the light sea quark mass corresponding to a pion mass at 330 MeV~\cite{Aoki:2010dy}. Both DWF and overlap fermions have good chiral symmetry and it is shown that $\Delta_{mix}$, which is a measure of mismatch in mixed action, 
is very small~\cite{Lujan:2012wg} and its effects on the nucleon properties have not been found to be discernible~\cite{Gong:2013vja}. 
Since the $O(m^2 a^2)$ discretization errors are found to be small in the study of the charmonium spectrum and $f_{D_s}$~\cite{Yang:2014sea}, this allows us to compute the spin for the charm quark on this lattice. In addition to the advantage in normalization as mentioned above,
the zero mode contributions to $2mP$ in the disconnected insertion (DI) and $q$ in Eq.~(\ref{eq:awi}), which are finite volume artifacts, cancel when the overlap operators are used for both of them.

    We adopt %TODO: cite 1511.09089 or other paper?
the sum method~\cite{Maiani:1987by,Deka:2008xr} where the ratio is taken and the insertion time of the $2mP$ quark loop and the topological charge $q$ is summed between $t_i +1 $ and $t_f -1$ where $t_i/t_f$ is the nucleon source/sink time. As a result, the ratio R($\Delta t, q^2$), where $\Delta t = t_f - t_i$, is linearly dependent on $\Delta t$ and the slope is the matrix element of the spin content from $2mP$ or $q$, 
\begin{equation} \label{ratio}
{\rm R}(\Delta t, q^2) {}_{\stackrel{\longrightarrow}{\Delta t \gg 1}} {\text const.} + \Delta t  \langle p' s \left| O \right| p s \rangle \frac{i | \vec{s} |} {\vec{q} \cdot \vec{s}},
 \end{equation}
from which we can obtain $m/m_N\, g_P (q^2)$ and $g_G(q^2)$ as functions of the momentum transfer squared $q^2$.

    As explained in detail~\cite{Gong:2013vja,Li:2010pw}, we adopt the $Z_3$-noise grid smeared source, with support on some uniformly spaced smeared grid points on a time slice, and low-mode substitution (LMS) which
improves the signal-to-noise ratio substantially, given the same computer resources. For the $24^3 \times 64$ lattice,
we place two smeared sources in each spatial direction, each with a Gaussian smearing radius of $\sim 4$ lattice spacing, and
have seen a gain of $\sim 6$ times of statistics in the effective nucleon mass as compared to that of one smeared source.
In view of the fact that the useful time window for the nucleon correlator $C(t)$ is less than 14 and we have 
$T=64$ slices in time, we put two grid sources at $t = 0$ and 32 simultaneously to gain more statistics from one inversion. Thus, our grid has the pattern of $(2,2,2,2)$ with two smeared grid sources in each of the space and time directions. 

Since both the strange and charm are from the disconnected insertion (DI), the calculation involves the 
product of the nucleon propagator and the quark loop. For the quark loop, we employ the low mode average (LMA) 
algorithm which entails an exact loop calculation for the low eigenmodes of the massive overlap fermion over all space time points on the lattice. On the other hand, the high modes of the quark loops are estimated with 4-D $Z(4)$ noise grid sources on $(4,4,4,2)$ grids and diluted for time slices and even-odd sites for a total of 4 inversions, each with one $Z(4)$ noise.  
   
The AWI splits the divergence of the axial current into two parts, i.e. $2mP$ and $q$, and the two parts reveal different aspects of the physics contribution. The pseudoscalar part is low-mode dominated for light quarks, where the first 200 pairs of overlap eigenvectors contribute more than 90\% of the vacuum value for the very light quarks and
$\sim 70\%$ for the strange~\cite{Gong:2013vja}.  The overlap Dirac operator $D_{ov}(x,y)$
in the definition of the topological term in Eq.~(\ref{top-charge}) is exponentially local with an
exponential falloff  of one lattice spacing~\cite{Draper:2005mh}. Thus, the anomaly part, being local, captures the high-mode contribution of
the divergence of the axial-vector current.

\begin{figure}[b]\centering
\subfigure
{\includegraphics[width=1.0\hsize]{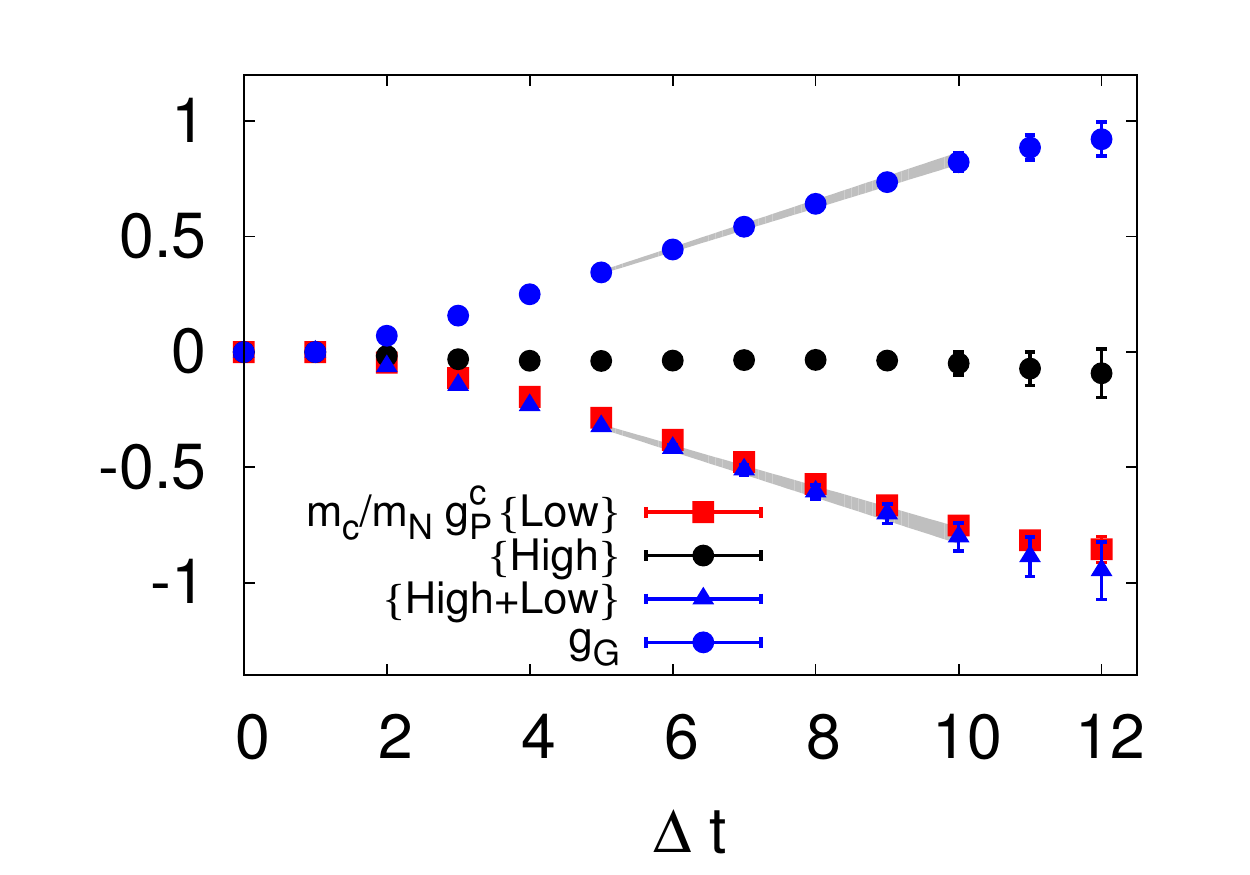}
%\label{fig:connected_insertion}
}
\hfill
\subfigure
{\includegraphics[width=1.05\hsize]{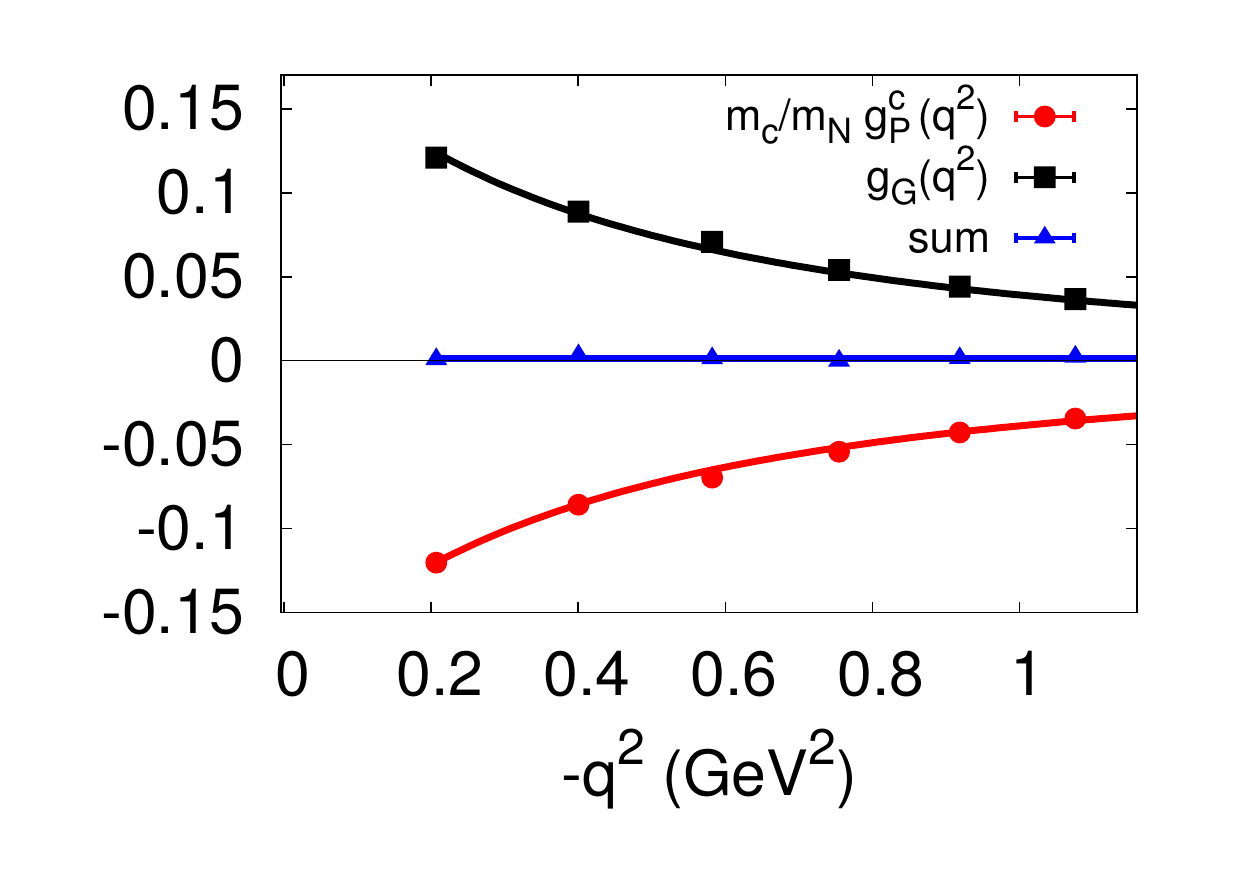}
 %\label{fig:disconnected_insertion}
 }
\caption{(Upper panel) The ratio of three-point and two-point correlators as a function of $\Delta t$ where the slopes are the
contributions from $2mP$ and $q$ at $|\vec{q}| = 2\pi/La$ in the DI for the charm quark in Eq.~(\ref{ratio}).
The red squares and the black points are the low and high-mode contributions respectively. The blue triangles with error band are the total. The valence 
  quark in the nucleon is the same as that of the light sea at $m_{\pi} = 330$ MeV. The similar ratio for the contribution from the
  topological charge $q$ is plotted as blue points whose slope gives $g_G(q^2)$.
  (Lower panel) The $2mP$  contribution $m/m_N g_P^c (q^2)$ and the anomaly contribution $g_G(q^2)$ are plotted as a function
  of $- q^2$.
  }
  \label{fig:charm}
\end{figure}

We first show the ratio in Eq.~(\ref{ratio}) for the charm quark as a function of $\Delta t$ for the case with lowest momentum
transfer, i.e. $|\vec{q}| = 2\pi/La = 0.469$ GeV (corresponding to $q^2 = - 0. 207 \,{\rm GeV}^2$) in the upper panel of Fig.~\ref{fig:charm}.  
The contributions from the low modes and high modes for $\frac{m_c}{m_N}\, g_P^c (q^2)$ at this $|\vec{q}|$,  
which are coded in the slopes, are shown separately. They are from the case where the valence quark in the nucleon and that of the light sea have the
same mass which correspond to $m_{\pi} = 330$ MeV.  It is clear from the upper panel of Fig.~\ref{fig:charm} that low modes dominate the
contributions. Even though the low modes contribute only $\sim 20\%$ in the charm quark loop itself~\cite{Gong:2013vja}, they become dominant when correlated with the nucleon. On the other hand, the $g_G(q^2)$ from the slope at this $|\vec{q}|$ is large and positive.
The errors for $\frac{m_c}{m_N}\, g_P^c$ and $g_G$ are 6\% and 4\% respectively.

\begin{figure}[hb]
\centering
\subfigure
{\includegraphics[width=1.0\hsize]{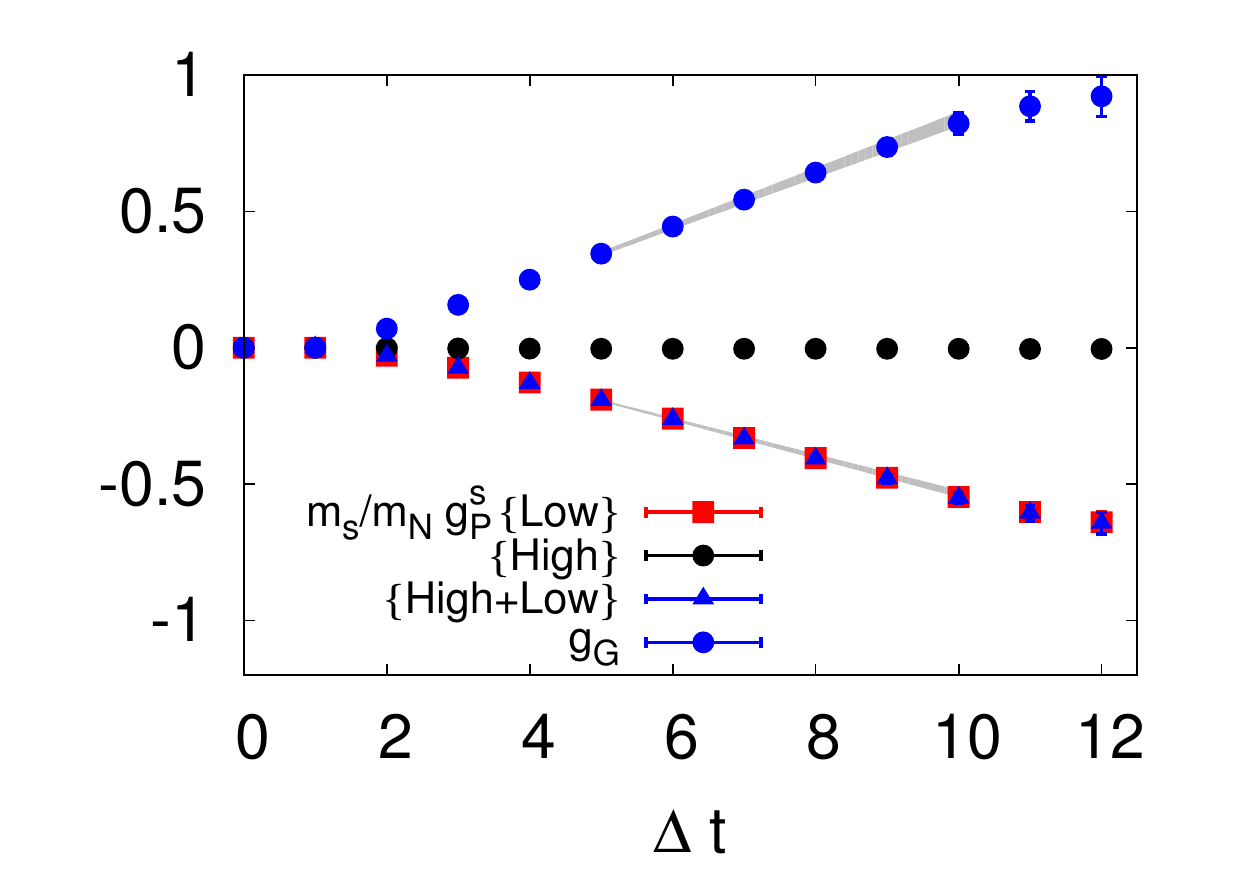}
%\label{fig:mass_ind}
}
\hfill
\subfigure
{\includegraphics[width=1.05\hsize]{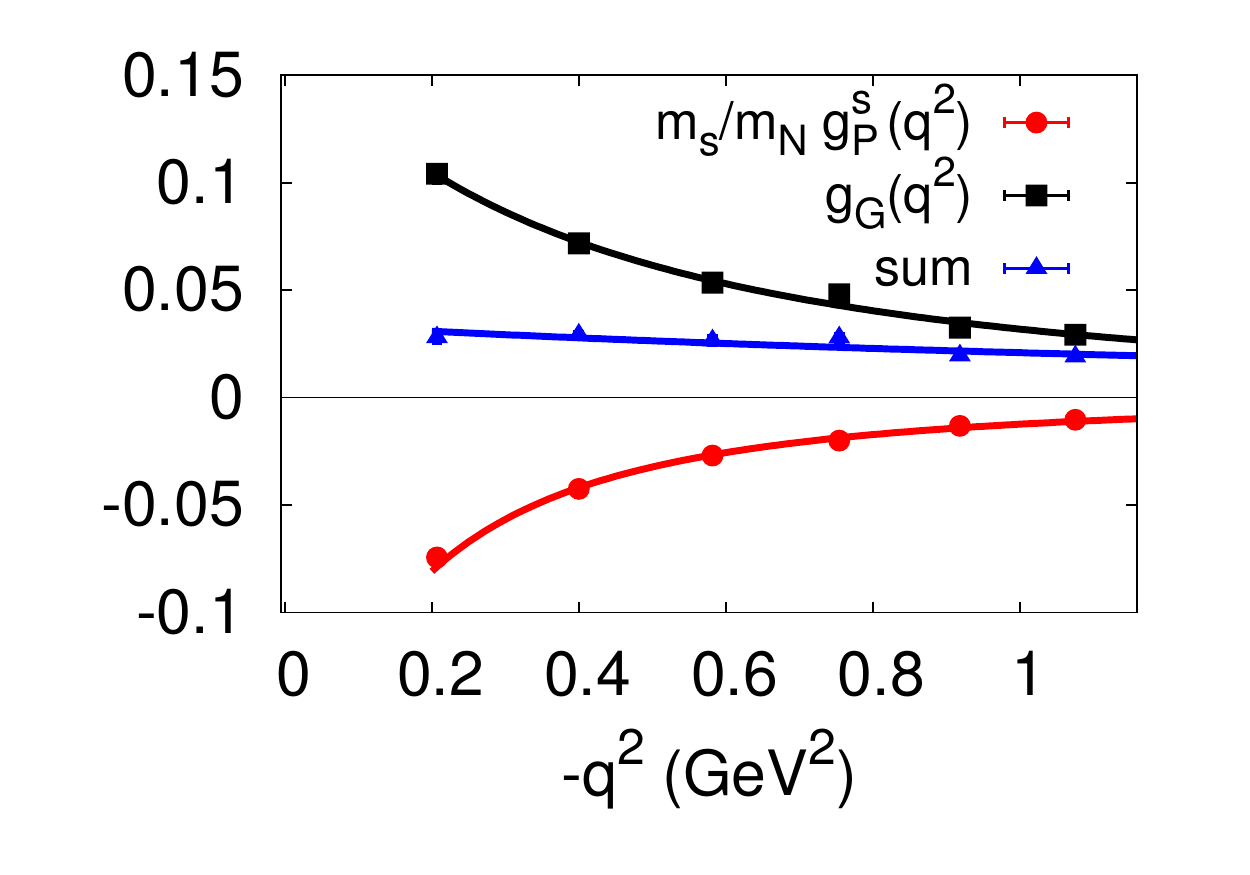}
 %\label{fig:charm_total}
 }
\caption{The same as in Fig.~\ref{fig:charm} but for the strange quark.}
 \label{fig:strange}
\end{figure}

In the lower panel of Fig.~\ref{fig:charm}, we give the results for the charm quark ($m_c^Ra = 0.73$) which is determined from
a global analysis of the charm mass~\cite{Yang:2014sea}. The pseudoscalar density term (red points) and the topological charge density term (black squares) are plotted as a function of $- q^2$. We see that the
pseudoscalar contribution is large, due to the large charm mass, and negative while the anomaly
is large and positive. The lines are fits with a dipole form just to guide the eye. When they are added together (blue triangles in the figure), they are very close to zero, with small statistical errors, over the whole range of $ - q^2$. Thus, when extrapolated to
$q^2 =0$ with a constant, we obtain $\Delta c + \Delta \bar{c} = -9.5(2.8)\times 10^{-4}$
at the unitary point.
When extrapolated to the physical pion mass, $\Delta c + \Delta \bar{c} = -2.7(2.8)\times 10^{-4}$,
which shows that the
charm hardly contributes anything, if at all, to the proton spin
due to the cancellation between the pseudoscalar term and the topological term.
It is known~\cite{Franz:2000ee} that the leading term in the heavy quark expansion of the quark loop of the pseudoscalar 
density, i.e.~$mP$, is the topological charge $\frac{i}{16 \pi^2}tr_c G_{\mu\nu} \tilde{G}_{\mu\nu}$ which cancels the contribution from the topological term in the AWI.
To the extent that the charm is heavy enough such that the $\mathcal{O}(1/m^2)$ correction is small, the present results of cancellation 
can be taken as a cross check of the validity of our numerical estimate
of the DI calculation of the quark loop as well as the anomaly contribution.
The mixing for the heavy quark loops from the other favors are also highly suppressed and negligible at the present stage.

Next, we consider the case with the strange quark ($m_sa = 0.063$) for this lattice, which is again determined from
the global fit for the strange quark mass based on fitting of $D_s$ and $D_s^*$~\cite{Yang:2014sea}.
Similarly to Fig.~\ref{fig:charm}, $\frac{m_s}{m_N}\, g_P^s(q^2)$ and $g_G(q^2)$ are plotted in Fig.~\ref{fig:strange} for
the unitary case where the valence quarks in the nucleon and the light sea quarks have the same mass at $m_{\pi } = 330$ MeV. 
We see in the upper panel that the low modes completely dominate the $2m_s P^s$ contribution as in the case of charm. The
anomaly is the same for all flavors. In the lower panel, it is shown that the contribution from $2m_sP^s$ is only slightly smaller
than that of the charm. This is due to the fact that even though the strange quark mass is about 12.5 times smaller than that of the 
charm~\cite{Yang:2014sea}, its pseudoscalar matrix element is much larger than that of the charm. 
Since the anomaly is the same for the strange and the charm, the sum of $\frac{m_s}{m_N}\, g_P^s(q^2)$ and $g_G(q^2)$, shown in
the lower panel, is slightly positive in the range of $- q^2$ as plotted.

Since our smallest $q^2 = - 0.207 {\rm GeV}^2$ is larger than $m_{\pi}^2$ which should be present as the pion pole on the right
hand side of the DI of AWI form factors to cancel that in the CI~\cite{Liu:1991ni,Liu:1995kb}, taking the $q^2 \rightarrow 0$ limit in Eq.~(\ref{eq:awi}) can lead to large systematic error.
In view of this, we calculated the unnormalized $g_A^L(q^2) = g_A(q^2)/\kappa_A$ and the induced pseudoscalar form factor $h_A^L(q^2)= h_A(q^2)/\kappa_A$ with the 3-point to 2-point correlator ratio $R(q_i, q_j)$~\cite{Deka:2013zha}
\begin{eqnarray}   \label{FF}
R(q_i, q_j, \Delta t) &{}_{\stackrel{\longrightarrow}{\Delta t \gg 1}}& {\text const.} + \Delta t \Big [ \frac{E_q + m_N}{2 E_q} \frac{g_A(q^2)}{\kappa_A} \delta_{ij} \nonumber \\
 &- &\frac{q_i q_j}{2E_q} \frac{h_A(q^2)}{\kappa_A}\Big ],
\end{eqnarray}
where $i$ and $j$ denote the directions of the axial current and the nucleon polarization. Here $g_A$ and $h_A$ are normalized
form factors. 
Sandwiching the AWI between the nucleon states with finite momentum transfer, one obtains
\begin{equation}  \label{AWIFF}
2m_N  g_A^{s (N)} (q^2) + q^2  h_A^{s (N)}(q^2) 
= 2m g_P^s (q^2) + 2m_N g_G(q^2).
\end{equation}
With 18 data points for $R(q_i, q_j)$ for different $q_i$ and 6 data points for $2m g_P^{s} (q^2)$ and $g_G(q^2)$ 
for 6 different $-q^2$,
we fit Eqs.~(\ref{FF}) and (\ref{AWIFF}) to obtain $g_A^{s(N)}(q^2)$ (including $g_A^{s(N)}(0)$), 
$h_A^{s(N)}(q^2)$, and $\kappa_A$. Since it is a global fit with all the $q^2$ data included,
this method does not require modeling the $q^2$ behavior with any assumed functional form.

\begin{figure}[htb]
\centering
{\includegraphics[width=1.0\hsize]{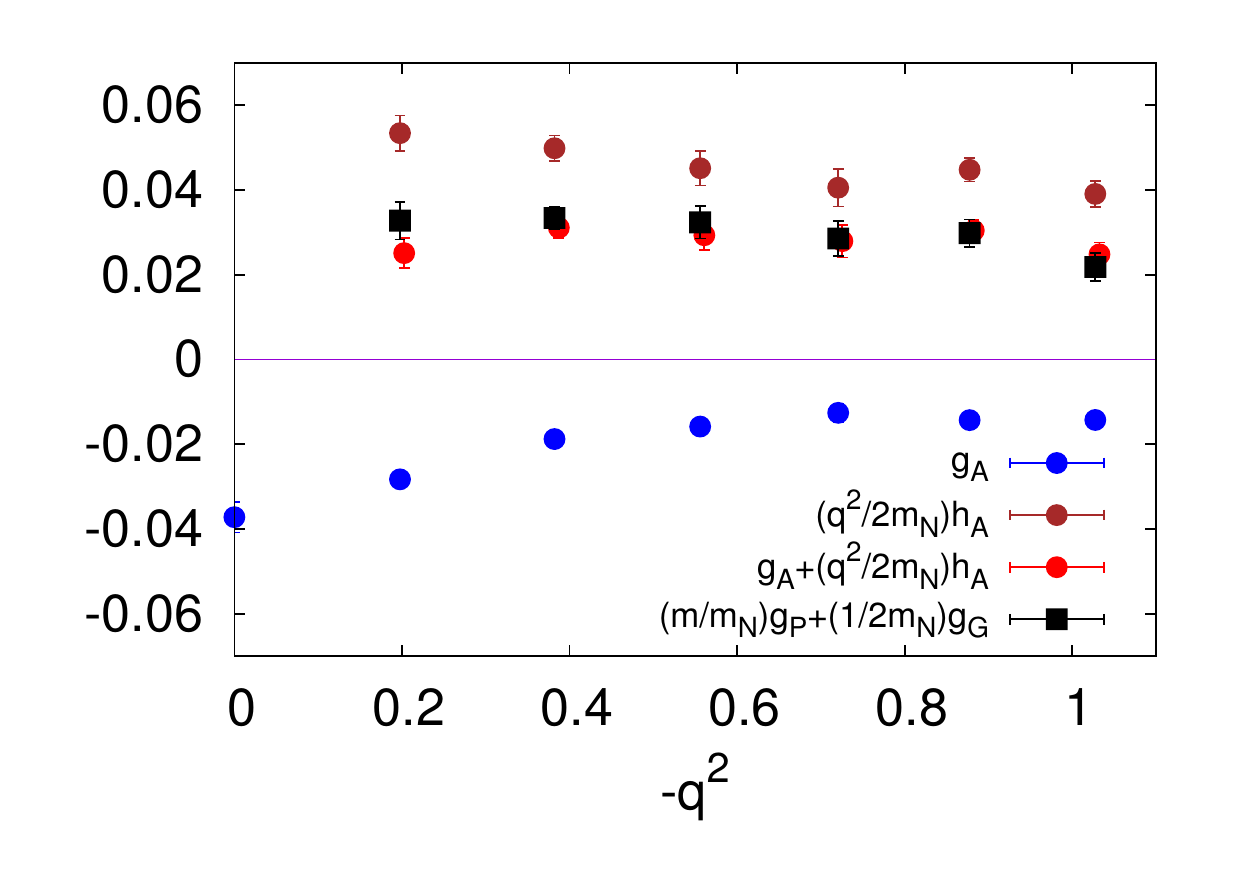}}
\caption{The $ - q^2$ dependence of the fitted normalized $g_A^{s}(q^2), \frac{q^2}{2m_N}h_A^{s}(q^2)$ and their sum in comparison
with $\frac{m}{m_N} g_P^{s}(q^2) + \frac{1}{2m_N}g_G(q^2)$. The latter is directly calculated. This is the case for the strange quark.} 
\label{fig:fit}
\end{figure}

The results for normalized $g_A^s (q^2)$, $h_A^{s}(q^2)$ are plotted in Fig.~\ref{fig:fit} as a function of $- q^2$. Also plotted is 
$g_A^{s(N)}(q^2) + \frac{q^2}{2m_N} h_A^{s(N)}(q^2)$ which is compared to 
$\frac{m}{m_N} g_P^{s} (q^2) + g_G(q^2)$ from the AWI in Eq.~(\ref{AWIFF}).
We see that the agreement is good for the range of $- q^2$ except for the last point at $- q^2 = 0.207 {\rm GeV}^2$ where there is a two-sigma difference.

\begin{figure}[h]
{\includegraphics[width=1.0\hsize]{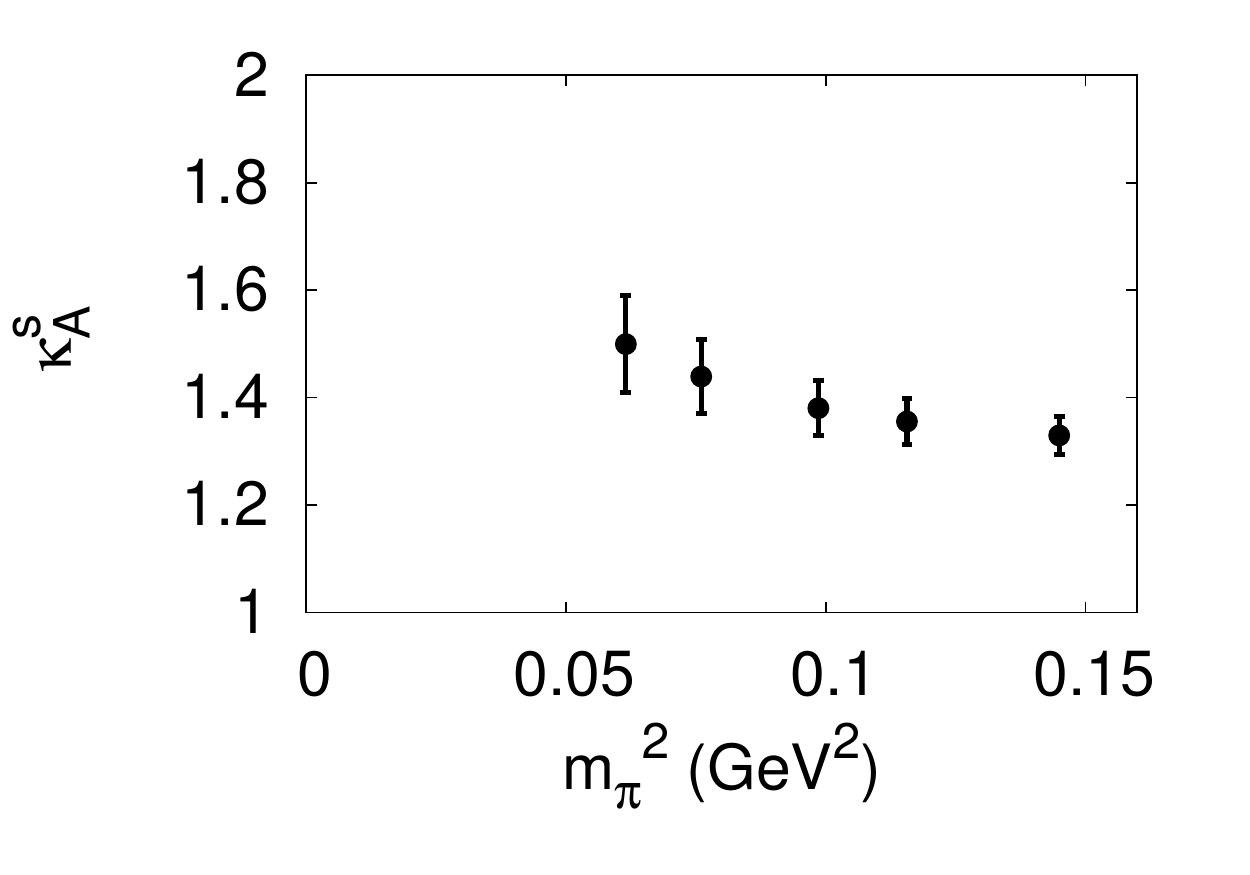}}
\caption{The normalization factor $\kappa_A$ as a function of valence $m_{\pi}^2$ while keeping the quark loop at the strange quark point.}
\label{fig:kappa}
\end{figure}

From the fit, we obtain $g_A^s = \Delta s + \Delta \bar{s} = -0.0372(36)$ \\
and \mbox{$\kappa_A = 1.36(4)$} at the unitary point where \mbox{$m_{\pi} = 330$ MeV. }
$\Delta s + \Delta \bar{s}$ and $\kappa_A$ have been calculated this way for several 
valence quark masses in the nucleon while keeping the quark loop at the strange quark point. 
The valence mass dependence of $\kappa_A$ is plotted in Fig.~\ref{fig:kappa}. We see that $\kappa_A$ is larger than 1, and becomes larger as the valence quark mass decreases.

\begin{figure}[h]
{\includegraphics[width=1.0\hsize]{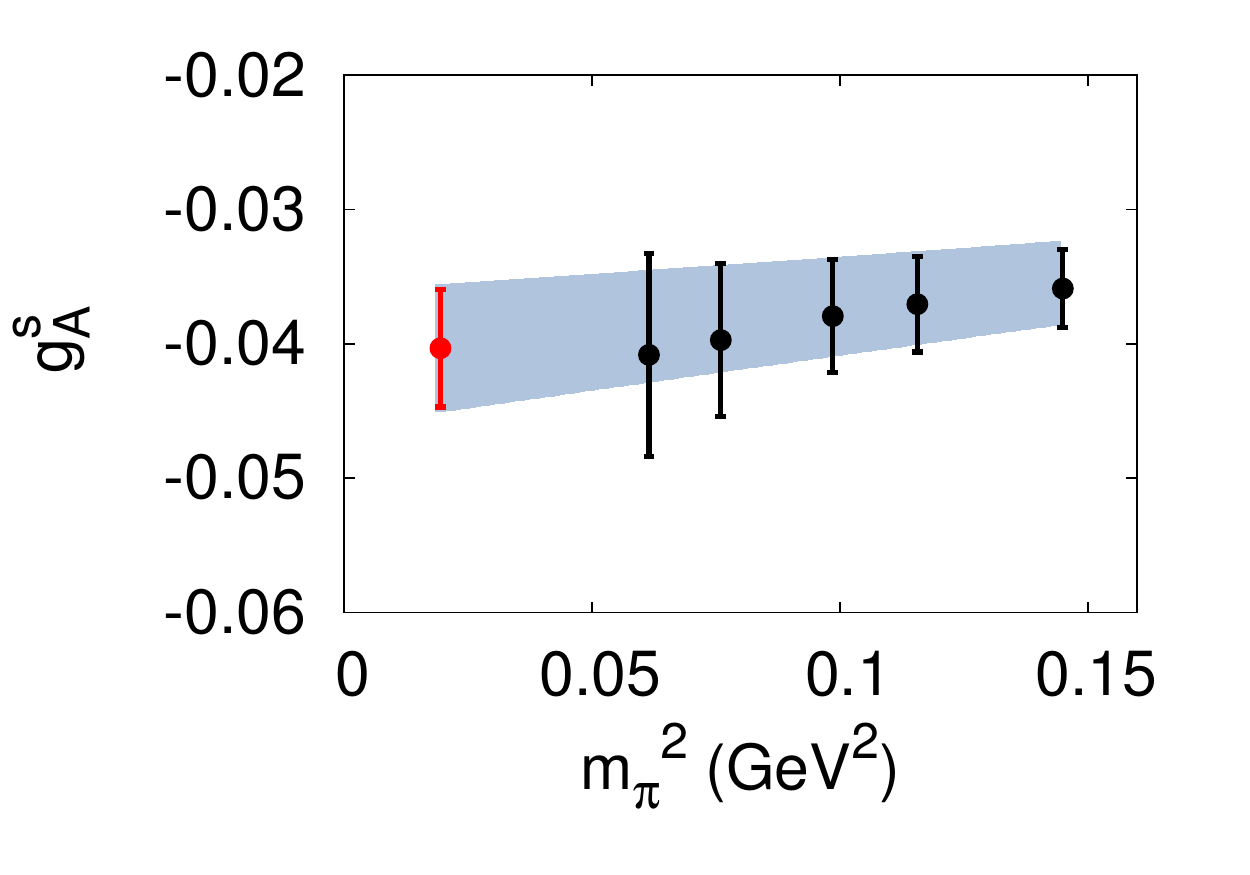}}
 \caption{Chiral extrapolation for the strange quark spin \mbox{$\Delta s + \Delta\bar{s}$} as a function of $m_{\pi}^2$.}
\label{fig:chiral}
\end{figure}

The chiral behavior of $\Delta s + \Delta \bar{s}$ is plotted in Fig.~\ref{fig:chiral}
as a function of $m_{\pi}^2$ according to the valence quark mass. 
We see that the results are fairly linear in $m_{\pi}^2$. Thus we
fit it linearly in $m_{\pi}^2$ with the form $A  + B(m_{\pi}^2 - m_{\pi, \rm{phys}}^2)$ where $m_{\pi, \rm{phys}}$ is 
the physical pion mass and obtain $\Delta s + \Delta \bar{s} = -0.0403(44)$ at the physical pion mass.
This is shown in Fig.~\ref{fig:chiral}. The uncertainty estimated through the variance from several different fits by adding a $m_{\pi}^2 \log(m_{\pi}^2/\Lambda^2))$ term, a $m_{\pi}^3$ term, or a $m_{\pi}^4$ term to the chiral extrapolation formula
gives a systematic error of 0.0013.

\begin{figure}[h]
{\includegraphics[width=1.0\hsize]{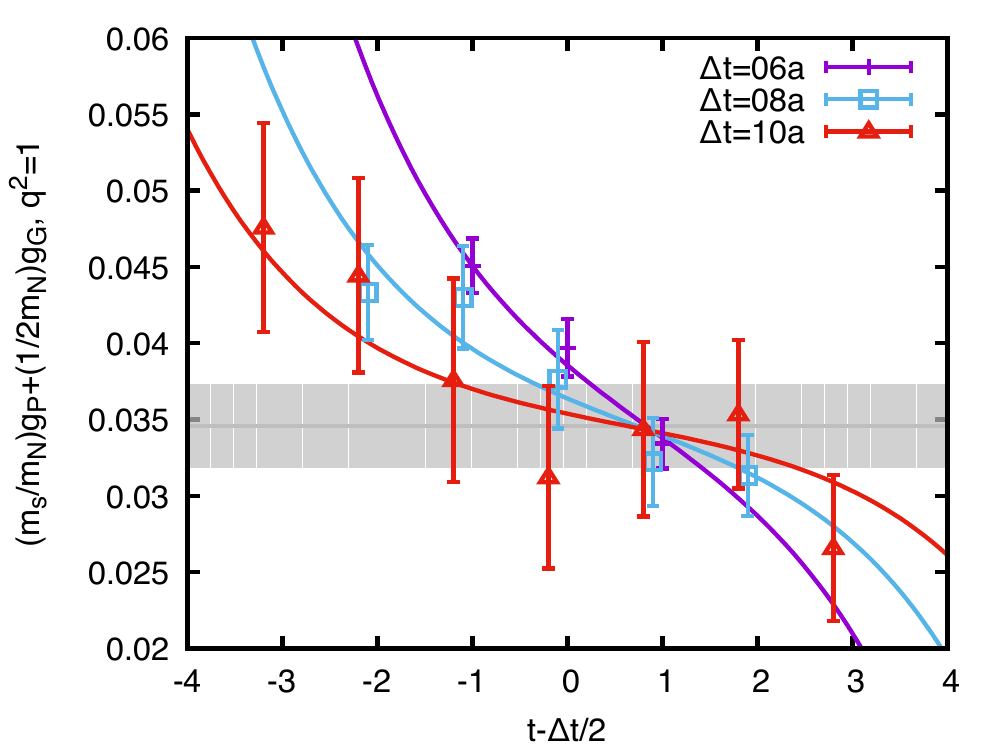}}
\caption{The 3-pt-to-2-pt ratio for $\frac{m_s}{m_N} g_P^s (q^2) + \frac{1}{2m_N} g_G (q^2)$ at the smallest 
$q^2 = -0.207 \rm{GeV}^2$ as a function of $t - t_f/2$. The separation $\Delta t = t_f - t_i = 6, 8, 10$ are shown with the data series. The lines on them are from two-state fit for the separate $\Delta t$. The grey band indicates the combined two-state and sum 
method fit.}
\label{fig:two-state-gP}
\end{figure}

In this work, we adopted the sum method to extract the matrix elements. To assess the excited state contamination,
we shall use the combined two-state fit with the sum method used in the calculation of the $\pi N$ and strange sigma 
terms~\cite{Yang:2015uis}, strange magnetic moment~\cite{Sufian:2016pex}, and glue spin~\cite{Yang:2016plb} for comparison 
for a few cases. We first plot  in Fig.~\ref{fig:two-state-gP} the un-summed ratios in Eq.~(\ref{ratio}) for 
$\frac{m_s}{m_N} g_P^s (q^2) + \frac{1}{2m_N} g_G (q^2)$ at the smallest 
$q^2 = -0.207 \rm{GeV}^2$ as a function of $t - t_f/2$ for time separations $\Delta t = 6, 8, 10$ between the source and the sink. A combined two-state and sum method fit with these data a value of 0.035(3) which is consistent within one sigma with the slope from the sum method which is 0.033(4). 

\begin{figure}[h]
{\includegraphics[width=1.0\hsize]{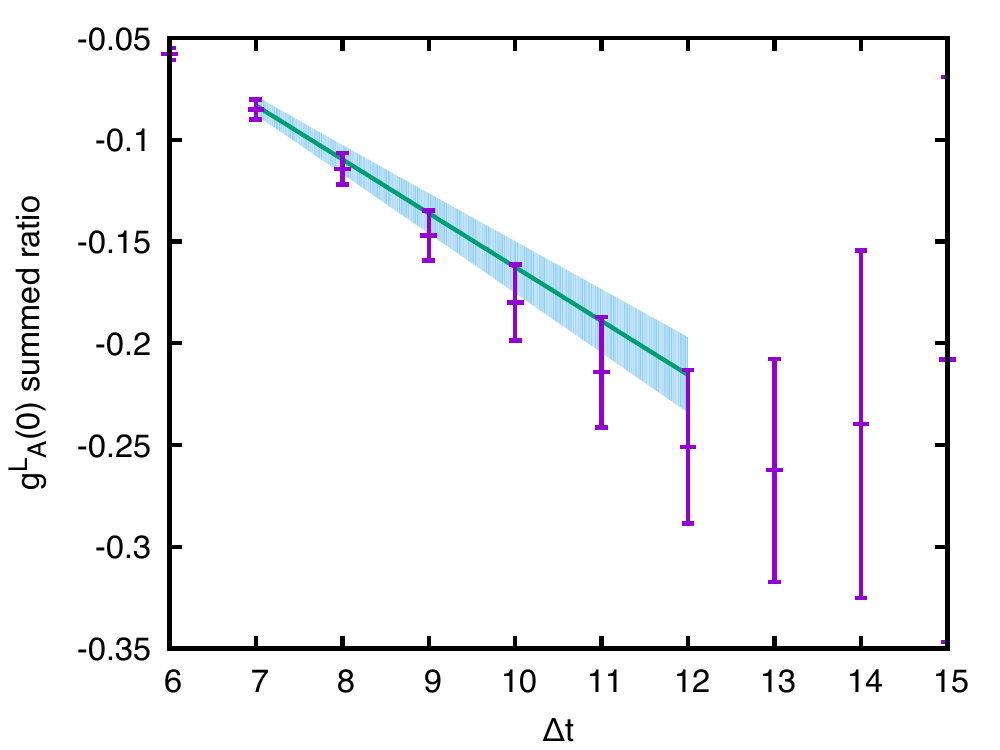}}
\caption{The summed ratio as a function of $\Delta t$ for the calculation of $g_A^L(0)$ which is extracted from the
slope as in Eq.~(\ref{ratio}).}
\label{fig:gA_sum}
\end{figure}

\begin{figure}[h]
{\includegraphics[width=1.0\hsize]{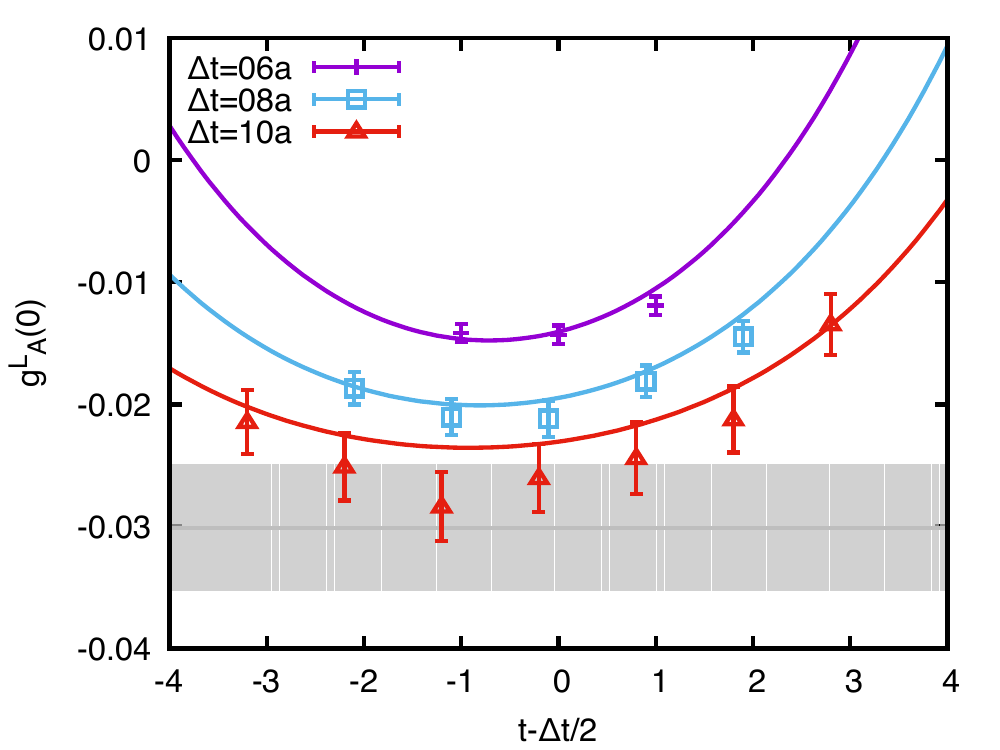}}
\caption{The same as in Fig.~\ref{fig:two-state-gP} for the bare $g_A^L(0)$.}
\label{fig:two-state-gA}
\end{figure}

Similarly, we have done the comparison for $g_A^L(0)$. Plotted in Fig.~\ref{fig:gA_sum} is the summed ratio 
of 3-pt-to-2-pt correlators as a unction of $\Delta t$ for the calculation of $g_A^L(0)$ which is extracted from the
slope as is from Eq.~(\ref{ratio}). At unitary point, we obtain $g_A^L(0) = -0.027(3)$. 
Also plotted in Fig.~\ref{fig:two-state-gA} are the un-summed ratios  for $g_A(0)$ as a function of $t - t_f/2$ for time separations $\Delta t = 6, 8, 10$ between the source and the sink. 
A combined two-state and sum method fit with these data a value of  -0.030(5).   While their errors touch, this is $\sim 10\%$ larger 
than that from the sum method fit. We shall take this 10\% difference as a systematic error of the present work.

The total systematic error contains the renormalization uncertainty $|\delta g_A^s| \sim 0.0066$, the uncertainty of
the chiral extrapolation of 0.0013, and uncertainty due to the excited state contamination of the sum method of 0.0040. We
sum them up quadratically and obtain an overall systematic error of 0.0078.

\begin{figure}[h]\centering
\includegraphics[width=1.0\hsize]{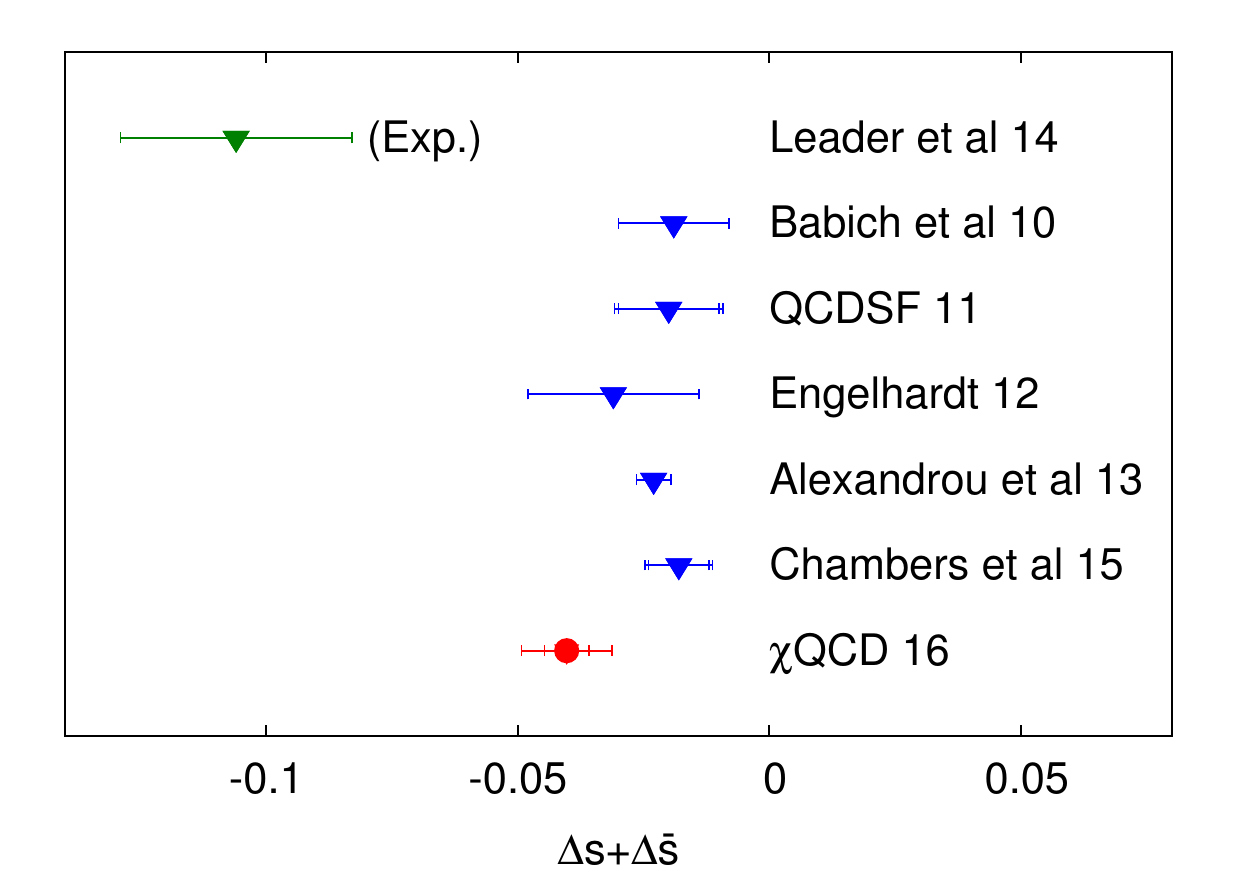}
\caption{A summary of the recent lattice QCD calculations of the strange quark spin $\Delta s + \Delta {\bar{s}}$ compared with the global fit of experiments. The blue triangles (color online) are lattice calculations from the axial vector current and the red circle (color online) is from the present work which uses the anomalous Ward identity.}
\label{fig:compare}
\end{figure}

We list our result in Fig.~\ref{fig:compare} together with other recent lattice results in comparison with the global fit of the DIS 
data~\cite{Leader:2014uua,Stamenov+Leader2015}. The blue triangles are lattice calculations of the axial vector current matrix element and the red circle is from the present work based on the anomalous Ward identity.

We see that our result, although still more than two sigmas smaller than the recent analysis of the DIS
data which finds the strange spin to be -0.106(23)~\cite{Stamenov+Leader2015}, is somewhat larger in magnitude than the other direct calculations of the axial-vector 
current~\cite{QCDSF:2011aa,Babich:2010at,Engelhardt:2012gd,Abdel-Rehim:2013wlz,Chambers:2015bka}.
This is mainly
due to the fact that the normalization factor $\kappa_A \sim 1.36$, which is required to have the AWI satisfied in our calculation, is 
larger than that for the isovector axial-vector current which is $1.10$ in our case.
Presumably, a similarly larger $\kappa_A$ exists for the other calculations using axial-vector currents which do not
satisfy the AWI, but has not been taken into account.

In summary, we have carried out a calculation of the strange and charm quark spin contributions 
to the spin of the nucleon with the help of the anomalous axial Ward identity. This is done with the
overlap fermion for the nucleon and the quark loop on $2+1$ flavor DWF configurations on
a $24^3 \times 64$ lattice with light sea quarks corresponding to $m_{\pi} = 330$ MeV. Since
the overlap fermion is used for the pseudoscalar term $2mP$ and the overlap Dirac operator is used
for the local topological term, the normalized AWI also holds for the renormalized 
AWI to two loop order.
For the charm quark, we find that the $2mP$ and the anomaly contributions almost cancel.
For the strange quark, the $2mP$ term is somewhat smaller than that of the charm.
Fitting the AWI at finite $q^2$ and the $g_A(q^2)$ and $h_A(q^2)$ form
factors, we obtain the normalized $g_A^s(0)$.
The normalization factor $\kappa_A \sim 1.36$ for the local axial-vector current is found to be larger than that
for the isovector axial-vector current,
 which implies that it is affected by a large cutoff effect presumably due to the triangle anomaly.
This will be clarified by future work using the conserved axial-vector current~\cite{Hasenfratz:1998ri} for the overlap fermion.
After chiral extrapolation to the physical pion mass for the nucleon, we obtain
$\Delta c + \Delta \bar{c} = -2.7(2.8)\times 10^{-4}$ which is consistent with zero,
and $\Delta s + \Delta \bar{s} = -0.0403(44)(78)$ which is smaller in magnitude than that from the latest analysis of DIS data~\cite{Leader:2014uua,Stamenov+Leader2015} by more than two sigmas.
We will check to see if this can be understood with lattices at the physical point.
In this work, we have identified the source for the negative spin contribution in the
disconnected insertion of the light quarks as due to the large and negative $2mP$ contribution which overcomes the positive anomaly contribution to give an overall negative $g_A^s(0)$. This is likely the cause for the smallness
of the net quark spin in the nucleon. We will confirm this later with results on the $u$ and $d$ quarks from both the disconnected
and connected insertions.

-------------------------------------

{\bf ACKNOWLEDGMENTS}

We thank RBC and UKQCD for sharing the DWF gauge configurations that we used in
the present work.  This work is supported in part by
the National Science Foundation of China (NSFC) under the project No. 11405178,
    the Youth Innovation Promotion Association of CAS (2015013),
    and the U.S. DOE Grant DE-SC0013065.
A.A. is supported in part by the National Science Foundation CAREER grant PHY-1151648.
This research used resources of the Oak Ridge Leadership Computing Facility at the Oak
Ridge National Laboratory, which is supported by the Office of Science of the U.S. Department of
Energy under Contract No. DE-AC05-00OR22725. This work also used the Extreme Science and Engineering 
Discovery Environment (XSEDE), which is supported by National Science Foundation grant number ACI-1053575.

\bibliographystyle{h-physrev3}
\bibliography{2014_strange_charm_quark_spin}

\end{document}